\documentclass[a4paper,11pt]{article}
\pdfoutput=1 % if your are submitting a pdflatex (i.e. if you have
\usepackage{jheppub} % for details on the use of the package, please
\usepackage[T1]{fontenc} % if needed
\usepackage{amsmath,amssymb,bm,epsfig,epsf,graphicx,float,url}
\usepackage{dsfont}
\usepackage{graphicx}
\usepackage{amsthm}
\usepackage{bbold}
\usepackage{graphicx}
\usepackage{caption}
\usepackage{subcaption}
\usepackage{wrapfig}
\usepackage{slashed}
\usepackage{feynmf}
\usepackage{hyperref}
\usepackage[english]{babel}
\usepackage{graphicx}
\usepackage{amssymb}
\usepackage{dsfont}
\usepackage{amsfonts}

\title{\boldmath Nonintegrability of NS5-like Interface \\ in $\mathcal{N}=4$ Supersymmetric Yang-Mills}

\author[a]{Miroslav Rap\v{c}\'{a}k}

\affiliation[a]{Perimeter Institute for Theoretical Physics,\\Waterloo, Ontario, Canada N2 2Y5}

\emailAdd{miroslav.rapcak@gmail.com}

\abstract{Four-dimensional $\mathcal{N}=4$ super Yang-Mills theory admits interfaces that preserve integrability of the theory. It was shown that addition of fundamental hypermultiplets living on a co-dimension one defect is an example of such an interface.  We consider NS5 interface as a different example of half-BPS defect in this theory and prove that integrability is generically broken already at one loop. We show that one-loop dilatation operator acting on boundary local operators does not lead to an integrable spin chain. The same is true also for D5-like and NS5-like boundaries.}

\begin{document}
\maketitle
\flushbottom

\section{Introduction}
\label{sec:intro}

Four dimensional $\mathcal{N}=4$ super Yang-Mills (SYM) has attracted a lot of attention within past few decades. Remarkably, this theory possesses some integrability features. The first evidence of  integrability \cite{Minahan:2002ve} came from the calculation of one-loop scaling dimensions of single trace operators composed of scalar fields of the theory in the large number of colors $N$ and the small 't Hooft coupling $\lambda=g^2N\ll1$ limit. The dilatation operator acting on such operators was shown to be equal to the Hamiltonian of the integrable spin chain with $SO(6)$ symmetry. This observation led to vast developments. See a comprehensive review \cite{Beisert:2010jr} and variety of recent papers on the topic.

$\mathcal{N}=4$ SYM with gauge group $U(N)$ appears as the effective low-energy theory of $N$ coincident D3-branes in type IIB string theory. One can consider the low-energy limits of more complicated brane configurations. For instance, we can ask the question if integrability is preserved once we include an D5-brane crossing our stack of D3-branes. We get a co-dimension one defect in our four dimensional theory with a fundamental hypermultiplet coupled to the bulk fields living on the interface \cite{DeWolfe1}. Instead of closed spin chains corresponding to the single trace operators, one obtains open spin chains with the fundamental defect fields forming boundaries of the spin chain. This system was shown to be integrable in \cite{DeWolfe:2004zt}.

Another natural possibility is to probe the integrability of spin chains associated to more general co-dimension one defects, constructed from D3-branes, D5-branes, and NS5-branes \cite{Gaiotto-Witten}. The simplest possibility is to include an NS5-brane crossing our stack of D3- branes. Perturbatively, each D3-brane is split into a left and a right halves. Bulk fields split into the left and right parts and they obey particular boundary conditions near the interface \cite{Hanany:1996ie}. Moreover, there is a 3d hypermultiplet in bifundamental representation of the two (left and right) gauge groups that couples to the two half-spaces.

Such brane construction is just a physical motivation for our calculations. We can also view it as a purely QFT system, where two $\mathcal{N}=4$ SYM theories with possibly different gauge groups and very natural boundary conditions (a particular combination of the Neumann and the Dirichlet boundary condition for different fields such that only half of supersymmetry is broken) are sewed together by bifundamental matter living at the interface.

We are interested in the scaling dimensions of single trace defect operators composed of scalar fields. There are three types of operators we can insert at the boundary. We have boundary limits of left and right bulk fields (those with non-vanishing boundary conditions) together with the sewing defect fields. Single trace operators are then composed of sequences of fields of the right part of the theory and sequences of the fields on the left sewed together by defect fields into the closed loop.

We will calculate the nontrivial part (term that leads to the operator mixing) of the one-loop dilatation operator of these operators and we prove that the theory is not integrable anymore. Even the subsector of operators consisting only of bulk fields living on one side of the defect breaks integrability due to the presence of non-trivial boundary conditions. The main argument is following. The form of the Hamiltoinan of the spin chain for bulk fields does not change (up to the overall factor of 2) if one includes the defect. On the other hand, symmetry of the spin chain is reduced from $SO(6)$ down to $SO(3)$ since half of the scalars vanish at the boundary. Integrable Hamiltonians with $SO(6)$ and $SO(3)$ symmetry have different forms and the integrability must be broken. The same argument also works in the case of D5-like and NS5-like boundary conditions.

This paper is organised as follows. In the section \ref{sec:interface}, we review bulk and defect content of the NS5-like interface together with boundary conditions imposed on the bulk fields. We also explicitly write down the interface action of this system together with propagators of all the fields in the theory. Section \ref{sec:bulk} sketches the calculation of the nontrivial part of the one-loop dilatation operator of boundary operators made of bulk fields only and argues that the integrability is not present. In the last section \ref{sec:boundary}, we show that integrability is broken even for NS5-like and D5-like boundary conditions.

\section{NS5-like interface}
\label{sec:interface}

In this section, we describe the boundary conditions for bulk fields, the boundary fields content, and the corresponding Lagrangian for the NS5-like interface in the $\mathcal{N}=4$ super Yang-Mills. NS5-like interface is a half-BPS defect in the theory corresponding to the low energy limit of two sets of infinite D3-branes attached to the NS5-brane from two sides. Detailed description of half-BPS boundary conditions and interfaces can be found in \cite{Gaiotto-Witten}.

Consider an NS5-brane spanning 012456 spacetime directions together with $N$ D3-branes in 0123 directions attached to the NS5-brane at $x^3=0$ both from the left and from the right.\footnote{We could possibly consider different numbers of D-branes attached from each side to the NS-brane.} In the low energy limit and far away from the boundary, we get four dimensional $\mathcal{N}=4$ SYM with $U(N)$ gauge group.  The bulk theory contains a gauge field $A_{\mu}$, six scalars $X^I$ in the \textbf{6} represetation of the R-symmetry group $SO(6)$, and a left-handed fermion in the \textbf{4} representation of the R-symmetry together with a right-handed fermion in the $\bar{\textbf{4}}$ representation of the R-symmetry.

The presence of the NS5-brane breaks Lorentz symmetry down to $SO(3)$ and R-symmetry down to $SO(3)\times SO(3) \sim SU(2)\times SU(2)$.  Under the broken R-symmetry group, the scalar fields decompose into two groups $X^I_V=(X^4,X^5,X^6)$ and $X^I_H=(X^7,X^8,X^9)$ that transform as (\textbf{3},\textbf{1}) and (\textbf{1},\textbf{3}) under this subgroup. Fermions decompose into four dimensional Majorana fermions transforming as (\textbf{2},\textbf{2}) under the broken R-symmetry.

Requiring D3-branes to end on the NS5-brane leads to boundary conditions imposed on bulk fields. In our special case, we get Neumann boundary condition for the gauge field
\begin{equation}
\label{eq:gauge}
F_{3\mu}=0.
\end{equation}
Half of the scalar fields $X^I_V$ that are part of the vector multiplet have to satisfy Neumann boundary conditions too. The rest of the scalars contained in the 3d hypermultiplet have to satisfy Dirichlet boundary conditions as shown in \cite{Gaiotto-Witten}
\begin{equation}
D_3X^I _V|_{x^3=0}=0,\qquad X^I _H|_{x^3=0}=0.
\end{equation}
In the case of fermions, correct boundary conditions are imposed simply by setting to zero half of the fermionic degrees of freedom at the boundary $\Psi^{a1}|_{x^3=0}=0$. The two indices correspond to the components of the two parts of the broken R-symmetry group.

The bulk action on each side of the interface is the standard $\mathcal{N}=4$ super Yang-Mills action, i.e.
\begin{eqnarray}\nonumber
S=\frac{1}{g^2}\int d^4x\ \mbox{Tr}&\bigg (&\frac{1}{2}F_{\mu \nu}F^{\mu \nu}+D _{\mu}X ^ID ^{\mu}X ^I-\frac{1}{2}[X ^I,X ^J]^2\\
&&-\ i\overline{\Psi}^a\slashed D \Psi^a +C_{Iab}\overline{\Psi}^a[X^I,\Psi^b]\ \bigg )
\end{eqnarray}
where the trace is performed with respect to flavor indices of the $U(N$) gauge group and $C_{Iab}$ are Clebsh-Gordan coefficients for the decomposition $\textbf{4}\times\textbf{4} \rightarrow \textbf{6}$.

Together with bulk fields described above, we have to deal with a defect hypermultiplet in the bifundamental representation of our two gauge groups. This hypermultiplet corresponds to the low energy limit of strings stretched between the two stacks of D3-branes on the left and on the right.

We can now follow the construction of \cite{DeWolfe1} to couple the defect hypermultiplet to the bulk theory. Using $\mathcal{N}=1$ superspace formalism in three dimensions the original four dimensional $\mathcal{N}=4$ vector multiplet can be represented by one 3d vector multiplet and three chiral multiplets. The defect $\mathcal{N}=4$ hypermultiplet in three dimensions can be written in terms of two complex $\mathcal{N}=1$ multiplets
\begin{equation}
Q^i=\Phi^i+\bar{\theta}\psi ^i+\frac{1}{2}\bar{\theta}\theta f^i
\end{equation}
transforming in the bi-fundamental representation\footnote{Let us stress that this is one of two small modifications of the construction in \cite{DeWolfe1}. In their case, they considered a fundamental representation instead.} with respect to the two (left and right) gauge groups
\begin{equation}
Q^i\rightarrow UQ^i\tilde{U}^\dagger
\end{equation}
where the tilded $\tilde{U}$ corresponds to the gauge transformation on the left of the boundary and $U$ corresponds to the right part. We have a pair of scalars $\Phi^i$, two-component fermions $\psi^i$, and auxiliary fields $f^i$. Coupling to the bulk gauge fields is given by the kinetic term of the 3d chiral multiplets
\begin{equation}
S_{kin}=\frac{1}{g^2}\int dx^3 d^2\theta\frac{1}{2}\mbox{Tr}\left [\overline{\nabla Q^i}\nabla Q^i\right ],
\end{equation}
where $\nabla$ is the covariantized 3d superspace derivative
\begin{eqnarray}
\label{covariant}
\nabla Q=\mathcal{D} Q-i\Gamma Q +iQ\tilde{\Gamma}.
\end{eqnarray}
In the expression above, $\Gamma$ and $\tilde{\Gamma}$ are spinor connection superfields obtained from restriction of the left/right bulk vector multiplet to the boundary. The spinor connections contain two terms, one is linear in $A_k$ and the other one in its superpartner $\Psi_a$.

To obtain an $\mathcal{N}=4$ supersymmetric theory, we have to add other couplings to the other bulk fields. We can now proceed in the same way as in \cite{DeWolfe1} by restricting a particular bulk superfield to the boundary, introducing a coupling of this restricted superfield to the defect degrees of freedom, and integrating out the auxiliary degrees of freedom. We could check explicitly that the resulting action has $N = 4$ supersymmetry, along the lines of the original calculation.

In this construction, one more modification is needed due to the non-trivial boundary conditions on the bulk fields.  When performing restriction of the bulk fields to the boundary, we need to set to zero all the terms containing fields with Dirichlet boundary conditions.

It is clear from the form of the defect covariant derivative \ref{covariant} that if $\Phi$ is a defect field and $X$ is a bulk field in the action constructed in \cite{DeWolfe1}, we need to do the replacement
\begin{eqnarray}
X\Phi\rightarrow X\Phi-\Phi \tilde{X}
\end{eqnarray}
where $X$ is now a bulk field on the right of the defect and $\tilde{X}$ is a bulk field on the left. Moreover, we have to cross off the terms containing bulk fields that vanish at the boundary. Explicitly, one gets\footnote{The $\delta (0)$ term will be explained momentarily. It comes from integrating out bulk auxiliary fields with nontrivial boundary conditions.}
\begin{eqnarray}\label{lagrangianNS5}
S_{kin}&=&\frac{1}{g^2}\int d^3 x\ \mbox{Tr}\left ((D^k\Phi^m)^{\dagger}D_k\Phi ^m-i\bar{\psi}^a\rho ^k D_k \psi ^a\right), \\
\nonumber
S_{yuk}&=&\frac{1}{g^2}\int d^3 x\ \mbox{Tr}\left (i\Phi^{\dagger m}\bar{\Psi}_{m0}\psi^0-i\Phi^{\dagger m}\bar{\psi}^0\tilde{\Psi}_{m0}-i\psi^{\dagger m}\Psi_{m0}\Phi^0+i\psi^{\dagger m}\Phi^0\tilde{\Psi}_{m0}\right )\\ \nonumber &&-\ \frac{1}{g^2}\int d^3 x\ \sigma_{mn}^I\ \mbox{Tr}\left (\bar{\psi}^mX_V^I\psi^n-\psi^m\tilde{X}_V^I\bar{\psi}^n\right ),\\ \nonumber
S_{pot}&=&\frac{1}{g^2}\int d^3 x\ \mbox{Tr}\left ( \Phi^{\dagger m}X^I_VX^I_V\Phi^m+ \Phi^{\dagger m}\Phi^m \tilde{X}^I_V\tilde{X}^I_V -  2\Phi^{\dagger m}X^I_V\Phi^m \tilde{X}^I_V \right )
\\&&+\ \frac{1}{g^2}\int d^3 x\ \sigma^I_{mn}\ \mbox{Tr}\left (\Phi^{\dagger m}(D_3X^I_H)\Phi^n-\Phi^{\dagger m}\Phi^n(D_3\tilde{X}^I_H)+\frac{1}{2}\delta (0)\mbox{Tr} (\Phi^{\dagger m}\Phi ^n)^2 \right ). \nonumber
\end{eqnarray}

There exists another way to construct the above action. We can view the four-dimensional bulk fields on one side of the interface as 3d fields with the gauge group $\hat{U}(N)$ of maps form $R^+$ to $U(N)$ with corresponding boundary condition at 0. In this way, we obtain two 3d vector multiplets (on each side of the theory) with the infinite dimensinal gauge group $\hat{U}(N)$ and two 3d hypermultiplets with the same infinite dimensional gauge group. Defect hypermultiplet can be the viewed as a multiplet on that trasnform in bifundamental representation after restriction to the boundary values of the two gauge groups $\hat{U}(N)$. We can then write down 3d $\mathcal{N}=4$ Lagrangian for these fields where the traces over the gauge groups $\hat{U}(N)$ are replaced by an integration over $R+$ and a trace over $U(N)$.

To construct propagators in the presence of these boundary conditions, we can use the standard mirror image trick.  One can simply verify that
\begin{equation}
\label{scal_prop}
\left \langle X^{IA}_{B} (x)X^{JC}_{D}(y)\right \rangle _\pm=\frac{1}{2}\frac{\delta ^{IJ}\delta ^{A}_D\delta _B^C}{4\pi^2}\left [\frac{1}{|x-y|^2}\pm \frac{1}{|x-\hat{y}|^2}\right ]
\end{equation}
is correct scalar propagator with Neumann ($+$) and Dirichlet ($-$) boundary conditions. In the expression above, we have denoted $\hat{x}=(\tilde{x},-x^3)$ for $x=(\tilde{x},x^3)$ and written explicitly the indices labeling different scalar fields $I$, $J$ and flavor indices of the gauge group $A$, $B$, $C$, $D$. The first term is the standard part that would be present even without the boundary, whereas the second term is its mirror counterpart which can be viewed as a propagation of a particle hitting the boundary. We will label this propagator (with the Kronecker delta omitted) $\Delta (x,y)$ for future convenience.

The propagator of the gauge field is a bit more complicated since we need to satisfy a less trivial condition \ref{eq:gauge}. Luckily, we have the choice of gauge in hand. Considering axial gauge $A_3=0$, we can completely forget about this component and the second term in the equation vanishes trivially. We are thus left with condition $\partial _3A_{\mu}=0$ that can be easily satisfied using the mirror trick. Propagator for a gauge field without a boundary in this gauge is
\begin{equation}
\Delta _{\mu \nu}(k)=-\frac{1}{2k^2}\left [g_{\mu \nu}-\frac{k_\mu k_\nu}{k_3^2} +\frac{\delta_{3\mu}k_\nu+\delta_{3\nu}k_\mu}{k_3}\right ]
\end{equation}
where we have now omitted flavor indices. Performing Fourier transform of this propagator, restricting to nontrivial components $\mu, \nu =0,1,2$ and using the mirror trick, we find
\begin{equation}
\Delta _{\mu \nu}(y,x)=
-\frac{1}{2}\int \frac{d^4k}{(2\pi)^4}\frac{1}{k^2}\left [g_{\mu \nu}-\frac{k_\mu k_\nu}{k_3^2}\right ]\left [e^{-ik\cdot (y-x)}+e^{-ik\cdot (y-\hat{x})}\right ].
\label{gauge_propagator}
\end{equation}

As already mentioned, half of the fermionic fields vanish at the boundary. Propagators of nonvanishing fields are then the same as in \cite{DeWolfe1}. For the rest of the fermions, we need to modify the propagator in the same way as in the previous cases of the Dirichlet boundary condition. As we will see later, both fermionic and gauge field propagators are not necessary in probing the integrability of the scalar sector of the defect fields since they give only a self energy contribution.

For completeness, let us also comment on the propagators of boundary fields. In the case of scalar fields, we have simply
\begin{equation}
\langle \Phi ^{iA}_{\bar{B}}(y) \Phi^{\dagger j\bar{C}}_{D}(x)\rangle=\frac{1}{2}\frac{\delta ^{ij}\delta ^A_D\delta ^{\bar{C}}_{\bar{B}}}{4\pi^2}
\frac{1}{|x-y|}
\end{equation}
where we have indicated all the indices to stress that they correspond to different gauge groups. Similarly as in the case of bulk fields, we will use $\mathcal{D}(y,x)$ for the part of the above propagator without the Kronecker deltas. Propagators of fermionic defect fields are the same as in the appendix of \cite{DeWolfe:2004zt} up to the overall factor of $1/2$ corresponding to different normalization of our fields.

As explained in \cite{DeWolfe1}, the term proportional to the delta function comes from integrating out bulk auxilary fields $F$ present in the superspace construction. To perform perturbative calculation we can go back to the action with $F$ not integrated out and include it to our calculations. In the original action (before integrating out $F$), we had a term of the form $\Phi^{\dagger}F\Phi$ that leads to an effective interaction between two $\Phi$'s. There is another interaction of this type corresponding to the propagation of $\partial _3 X_H^I$. Adding those two pieces together, we  can effectively describe the propagation of both $\partial _3 X_H^I$ and $F$ using a single diagram, where the propagator of the intermediating particle has the following form
\begin{equation}
\Delta _{\partial _3X_H+F}(y,x)=\frac{1}{\pi ^2}\frac{1}{(y-x)^4}.
\label{aux_propagator}
\end{equation}
In our case, we had to take into account nontrivial boundary conditions on the the fields $\partial _3 X_H^I$ and $F$. The propagator of $\partial _3 X_H^I$ can be obtained from the propagator of  $X_H^I$ performing two derivatives. On the other hand, $F$ has to satisfy Neumann boundary conditions as can be seen from the equations of motion for this field. Putting these two facts together, using the standard mirror trick, and the argument from \cite{DeWolfe1}, we get the combined propagator above \ref{aux_propagator}.

\section{Boundary operators made of bulk fields}
\label{sec:bulk}

We would like to investigate integrability of the dilatation operator acting on local operators. Let us thus split 4d coordinates as $x=(\tilde{x},x_3)$ where the boundary is located at $x_3=0$. For fields inserted at the boundary, we will write just $\tilde{x}$ as an argument to simplify our notation. In this section, we will see that one-loop dilatation operator is not integrable even if we restrict to the boundary limit of the bulk scalar fields
\begin{equation}
\label{singletrace}
\mbox{Tr}\left [X_V^{I_1}\dots X _V^{I_n}\right ](\tilde{x})
\end{equation}
where $I_k=1,2,3$. Other scalar fields vanish at the boundary due to the boundary conditions and only their normal derivatives
will appear in boundary local operators.\footnote{In the case of D5-brane like interface, we just have to replace $V\leftrightarrow H$ in the subindex.} Alternatively, we could consider fields $\tilde{X}_k$ living on the other half of the spacetime and we would get the same result. Restricting to gauge invariant, single trace operators, the two sectors do not mix unless we include also sewing defect fields. This fact will be discussed in more detail later.

\begin{figure}[h!]
  \centering
      \includegraphics[width=0.9\textwidth]{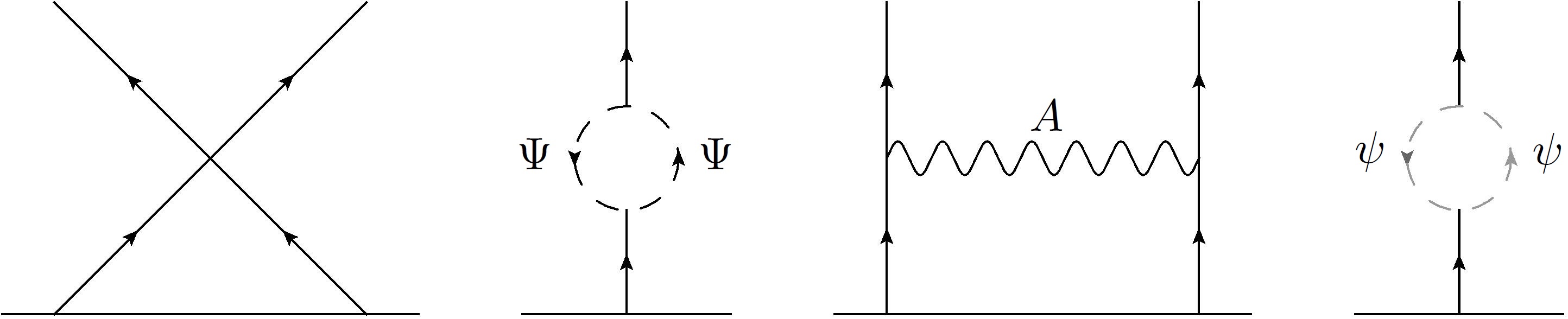}
  \caption{Diagrams contributing to the one-loop dilatation operator of the single trace operator \ref{singletrace}. Note the new contribution due to the presence of defect fields (labeled by gray lines).}
\label{fig:feynman}
\end{figure}

As in the original computation in \cite{Minahan:2002ve}\footnote{See also references therein or a nice review \cite{Beisert:2010jr}.} the only nontrivial contribution that results to operator mixing comes from the four-vertex of scalar fields. The other self-energy one-loop contributions contain bulk fermion loop and gauge boson loop as in the original story. Here, we have to include also the self-energy contribution coming from the loop of the boundary fermions $\psi$ (shown together with all the other contributions in the figure \ref{fig:feynman}) as clear from the Lagrangian \ref{lagrangianNS5}. At first sight it is not clear that this diagram does not change the index of $X_V^{I_k}$ or even mix the two bulk sectors (i.e. produces $\tilde{X}_V^{I_k}$ in the spin chain picture) due to the presence of Pauli matrices in the corresponding term.  Luckily, the Pauli matrices appear in a pair (we have two vertices of the same type) and we get a contribution proportional to
\begin{equation}
\sigma ^I_{mn}\sigma^J_{mn}\sim \delta ^{IJ}.
\end{equation}
We see that this diagram does not change the index $I_k$. On the other hand, mixing of the two sectors is suppressed in the 't Hooft limit since the corresponding diagram do not produce any power of $N$.

Let us argue that operators \ref{singletrace} form a closed subsector and they do not mix with other fields at one-loop. It is indeed true if we consider only bulk diagrams. We can thus restrict to processes containing at least one sewing defect field. Single field $X_V^{I_k}$ cannot change into a different field since there is no diagram that would allow it. Splitting of a single $X_V^{I_k}$ into two fields (possibly two boundary scalars) is also not present in the large $N$ limit. The last case we need to discuss is the situation when two $X$'s combine and produce some other combination of two operators. Two boundary scalar fields $\Psi$ cannot be produced (see the $S_{pot}$ part of the defect action) since the operators with different classical scaling dimensions do not mix. Another option is a production of two boundary fermions $\Psi$ using two 3-vertices in the second line of $S_{yuk}$. This process is not present in the $N \rightarrow \infty$ limit since the amplitude is proportional to a single power of $N$.

The contribution to the two-point function from the only nontrivial diagram mentioned above is
\begin{equation}
\big \langle \big (X ^{I_k}(x_1)X^{I_{k+1}}(x_2)\big )^A_C\left (-\frac{g^2}{2} \int_{z^3>0}  d^4z\sum _{I,J}\mbox{Tr}\big [ X ^I, X ^J \big ]^2(z) \right )\big (X^{J_{k+1}}X ^{J_k}\big )_{A'}^{C'}(0)\big \rangle.
\end{equation}
The structure of flavor indices and their contractions gives terms proportional to $N^2$ and $N$. In the planar limit, we can consider only the first case giving us the following index structure
\begin{eqnarray}\nonumber
\label{ref1}
=&g^2N^2\delta ^A_{A'}\delta ^{C'}_C\int_{z^3>0}  d^4z\ \Delta_+ (x_1,z)\Delta_+ (x_2,z)\Delta_+ ^2(z,0) \\
&\cdot
\left (2\delta  ^{J_{k}}_{I_k}\delta _{I_{k+1}}^{J_{k+1}}+2\delta _{I_{k}I_{k+1}}\delta ^{J_{k}J_{k+1}}-4\delta ^{J_{k+1}}_{I_k}\delta _{I_{k+1}}^{J_{k}}\right ).
\end{eqnarray}
Using the expression for the scalar propagator \ref{scal_prop}, introducing a sharp cut-off $|z|>M$ followed by an integration gives us
\begin{eqnarray}
\label{ref2}
\approx  g^2N^2\delta ^A_{A'}\delta ^{C'}_C\frac{1}{(4\pi^2)^3}\frac{1}{x_1^2}\frac{1}{x_2^2}\mbox{ln}\left (M ^2|x_1|^2\right )\left (\delta  ^{J_{k}}_{I_k}\delta _{I_{k+1}}^{J_{k+1}}+\delta _{I_{k}I_{k+1}}\delta ^{J_{k}J_{k+1}}-2\delta ^{J_{k+1}}_{I_k}\delta _{I_{k+1}}^{J_{k}}\right )
\end{eqnarray}
where we have assumed $|x_1|^2<|x_2|^2$ without loss of generality.
Comparing to the case of free propagation
\begin{equation}
\langle (X _{I_k}(x_1)X_{I_{k+1}}(x_2))^A_C(X^{J_{k+1}}X ^{J_k})^{C'}_{A'}(0)\rangle =\frac{N}{(4\pi^2)^2}\delta ^A_{A'} \delta_C^{C'}\frac{1}{x_1^2}\frac{1}{x_2^2}\delta _{I_k}^{J_k}\delta_{I_{k+1}^{J_{k+1}}}
\end{equation}
where we again consider only the contraction corresponding to the leading order in $N$, and substituting into the Callan-Symanzik equation,  we get a one loop scaling dimensions matrix
\begin{equation}
\Gamma _{bulk}=\frac{\lambda}{8\pi ^2}\sum _{l+1}^L\left [1+C+K_{l,l+1}-2P_{l,l+1}\right  ] \label{bulk_nontrivial}
\end{equation}
where the operators $1$, $K$, and $P$ correspond to the 3 combinations of delta functions in \ref{ref2} as in \cite{Minahan:2002ve} and $L$ is the length of our spin chain (the number of scalar fields in the single trace operator). The constant factor $C$ corresponds to the other diagrams that do not influence integrability. As such they are not relevant for our discussion of integrability of the interface.

Note also that this operator is (possibly up to a constant additive factors) twice the dilatation operator of the theory without defect. In fact, we could have guessed this result immediately simply counting factors of 2. Near the boundary, propagator of bulk fields is twice the propagator without the presence of boundary. This can be viewed as two contributions coming from the direct propagation and the propagation bouncing the boundary. These two factors equal for boundary to boundary propagations. In our case, we have two particles propagating to the interaction point and two particles propagating from this point to the final point giving a factor of 2$^4$. Integration over the position of the interaction is only over half of the space giving a relative factor of 1/2. Comparing this to free propagation of the two fields that gives factor of 2$^2$, we get the desired 2. Apart from the propagator and range of the integration, nothing changes in this case and especially the index structure remains the same.

Let us also mention a method of differential regularization \cite{Freedman:1991tk} that is useful in avoiding integrations in our calculations.  To get rid of the integration, let us ampute external legs (i.e. apply $-2\Box$ on $x_i$'s of the external lines) in the corresponding Feynmann diagrams. At tree level, we get for example for the boundary to bulk propagation of the scalar field
\begin{equation}
4\Box_{x_1} \Box_{x_2} \langle (X(x_1)X(x_2))(XX)(0)\rangle =4N\delta (x_1)\delta (x_2)
\end{equation}
where we omitted all the unnecessary indices. The factor of 4 is coming from the two mirror terms in the propagator in the presence of the boundary. The integral $I$ in \ref{ref1} gives
\begin{eqnarray}\nonumber
4\Box_{x_1} \Box_{x_2} I &=&\frac{2}{(4\pi^2)^2} \int_{z^3>0} \frac{d^4z}{z^4}\big (\delta (x_1-z)+\delta (x_1-\hat z)\big )\big (\delta (x_2-z)+\delta (x_2-\hat z)\big)\\
&=& \frac{2}{(4\pi^2)^2}\frac{1}{x_1^4}\delta (x_1-x_2).
\end{eqnarray}
Now, we can perform a regularization using the identity
\begin{equation}
\frac{1}{x_1^4}=-\frac{1}{4}\Box \frac{\mbox{ln}x_1^2\Lambda ^2}{x_1^2}.
\end{equation}
Substituting into the above expression and then to the Callan-Symanzik equation and using $\Box 1/x^2=-4\pi^2\delta (x)$, we get the same result as before \ref{bulk_nontrivial}.

It is now clear that the integrability is not preserved by the NS5-like interface. Spin chains we are considering here have $SO(3)$ symmetry corresponding to the R-symmetry rotation of the three scalars that do not vanish at the boundary. Demanding integrability of a spin chain constraints the form of Hamiltonians composed of $1,K$ and $P$ operators considerably. It forces the ratio of the coefficients standing in front of $P$ and $K$ operators to be $(n-2)/2$ for the model with $SO(n)$ symmetry.\footnote{Another possibility is that the coefficient in front of $K$ vanishes.} The only parameters that can be adjusted in order to get an integrable model are the additive constant and the overall multiplicative factor.  In our case, the ratio is the same that we had in the $SO(6)$ case of the theory with no defect but the symmetry group has changed. We have to conclude that integrability is broken even for the subsector containing only boundary values of bulk fields.

\section{D5-like and NS5-like boundaries}
\label{sec:boundary}

In this section, we are going to comment on integrability of NS5-like and D5-like boundary. Let me start with the NS5-like boundary whose non-integrability is just straightforward consequence of the above calculation. To get an NS5-like boundary, we restrict ourselves to the one half-space and we forget about all the boundary fields. We are thus left with a gauge field with Neumann boundary conditions, half of the scalars satisfying Neumann boundary condition and half of them satisfying Dirichlet boundary condition.

Calculation from the section \ref{sec:bulk} is still valid with the only exception that the last self-energy contribution from the picture \ref{fig:feynman} is not present. We are thus left with the same situation as before and we have to conclude that integrability is broken even in this case.

In the case of D5-like boundary, the gauge field satisfy Dirichlet boundary condition
\begin{eqnarray}
F_{\mu\nu}=0 \qquad \mbox{for } \mu, \nu \neq 3.
\end{eqnarray}
Moreover, boundary conditions of scalars are exchanged. Since the gauge field satisfy Dirichlet boundary condition, there is no obvious way to couple boundary degrees of freedom to the bulk theory.

Situation in the presence of the D5-like boundary is the same as in the case of the NS5-like boundary but now with the other triple of scalar fields. Calculation from the section \ref{sec:bulk} is again valid and integrability is thus spoiled also in this case. Note also that propagator of the gauge field is different and the self-energy contribution coming from the third diagram in \ref{fig:feynman} might be different.

\section{Conclusion}
\label{sec:discussion}

In this short note, we probed integrability of the NS5-like interface in the $\mathcal{N}=4$ SYM theory. We wrote down the Lagrangian of the theory and all the relevant propagators. Unlike the case of coupling fundamental hypermultiplet localized on the 3d surface, we have to deal with non-trivial boundary conditions imposed on bulk fields that break integrability by themselves. Integrability is broken even in the subsector of boundary values  of bulk fields. Since the breakdown of integrability is result of non-trivial boundary conditions, integrability is broken even if we consider D5-like or NS5-like boundary condition.

One could in principle finish calculations of one-loop scaling scaling dimensions of boundary fields calculating contributions from other diagrams that do not lead to the operator mixing. We could also consider more general single trace operators composed of both types of defect fields (boundary values of bulk fields, sewing fields living on the defect). Since we have seen that operators considered here form closed subsector under the operator mixing (at one loop), this would not save the integrability. One can also calculate scaling dimensions of single trace operators going to higher loops but this analysis does not seem to be interesting since we are dealing with a non-integrable system. Another possible direction is the question of integrability of different interfaces and possibly other defects in the theory. The simplest extension would be introduction of a general CFT living on the defect.

\acknowledgments
I am grateful to Davide Gaiotto for suggesting this project, reading the manuscript, and interesting discussions and explanations. I would like to thank Lakshya Bhardwaj, Nadav Drukker, Dalimil Maz\'{a}\v{c}, and Pedro Vieira for discussions.
The work of the author was supported by the Perimeter Institute for Theoretical Physics. Research at
Perimeter Institute is supported by the Government of Canada through Industry Canada
and by the Province of Ontario through the Ministry of Economic Development and Innovation.

% We however suggest t\\\/o always provide author, title and journal data:

\end{document}